\documentclass[aps,pre,showpacs,twocolumn,superscriptaddress,letter]{revtex4-1}
\usepackage{epsfig}
\usepackage{amsmath,amssymb,wasysym,epsfig,capt-of,ifthen,calc}
\usepackage{bbm} 
\usepackage{latexsym} 
\usepackage{setspace} 
\usepackage{array} 

\usepackage{delarray} 
\usepackage{afterpage} \usepackage{graphicx}
\usepackage{dcolumn}
\usepackage{bm}
\usepackage{float} \usepackage{supertabular} \usepackage{longtable}
\newcommand{\be}{\begin{equation}} \newcommand{\ee}{\end{equation}}
\newcommand{\bea}{\begin{eqnarray}} \newcommand{\eea}{\end{eqnarray}}

\usepackage{amsmath}\bibstyle{apsrev}
\begin{document} \title{ The Interacting Branching Process as a Simple Model of Innovation}

\author{Vishal Sood} \affiliation{Complexity Science Group, University
  of Calgary, Calgary, Canada} \affiliation{Center for Models of Life,
  Niels Bohr Institute, Copenhagen, Denmark}\affiliation{Niels Bohr
  International Academy, Blegdamsvej 17, DK-2100, Copenhagen, Denmark}
 \author{Myl\'ene Mathieu}
\affiliation{\'Ecole Normale Sup\'erieure de Lyon, Universit\'e Claude
  Bernard Lyon, France} \author{Amer Shreim} \affiliation{Complexity
  Science Group, University of Calgary, Calgary, Canada} \author{Peter
  Grassberger} \affiliation{Complexity Science Group, University of
  Calgary, Calgary, Canada} \author{Maya Paczuski}
\affiliation{Complexity Science Group, University of Calgary, Calgary,
  Canada}
\begin{abstract} 
  We describe innovation in terms of a generalized branching
  process.  Each new invention pairs with any existing one to produce a
  number of offspring, which is Poisson distributed with mean $p$.
  Existing inventions die with probability $p/\tau$ at each
  generation.  In contrast to mean field results, no phase
  transition occurs; the chance for survival is finite for all $p>0$.  
  For $\tau=\infty$, surviving processes
  exhibit a bottleneck before exploding super-exponentially -- a growth consistent with a law of accelerating returns.  This behavior persists
  for finite $\tau$.  We analyze, in detail, the asymptotic behavior  as $p \to 0$.
\end{abstract} \pacs{02.70.Uu, 05.10.Ln, 87.10.+e, 89.75.Fb, 89.75.Hc}
\maketitle

Networks of elements creating 
new elements can be found in diverse conditions -- from the origin
of life to artistic expression. As such, innovation and discovery are
intrinsic to life as well as human experience.  Economic,
technological or social innovations include introducing a new good,
method of production, form of governance,  etc.  Clearly,
innovations do not happen in a vacuum.  They form contingent and
interconnected webs where existing discoveries foster the creation of
new ones -- leading to a network of self-perpetuating, autocatalytic
activity. Landmark innovations ignite radical change --
avalanches of new discoveries that were previously
unthinkable~\cite{Burke}.  These include, for instance, the World Wide
Web,  or the
bursts of creativity associated with the Renaissance.  But innovation
is not restricted to human society. It is an archetype for any
co-evolutionary dynamics. Indeed the intermittent, bursty pattern of
innovation in human history resembles punctuated equilibrium observed
in biological evolution -- where episodes, such as the Cambrian
explosion, dominate the history of life's diversity~\cite{sneppen}.

Previous theoretical approaches to this phenomenon have concentrated on
mean field models, where explosions of innovation were attributed to a
phase transition~\cite{Farmer86,Hanel05,Hanel07}. This transition separates 
a regime where activity
always dies out after giving rise to only a few elements from a phase
where large numbers of elements (or inventions) can come forth.  The
phase transition depends both on the number of initial -- or primeval -- 
inventions and the ``mating'' probability to create new elements from
interactions between existing ones. 

Here we take a different approach that explicitly considers
fluctuations in a microscopic model of innovation.  Pairs of
inventions mate to create new ones.  The relevant criterion is the
chance for the process to survive forever, or, in other words to
escape extinction.  In contrast to conclusions based on mean field
arguments, our analysis establishes that fluctuations always lead to a
finite probability to escape extinction for any mating probability
$p\neq 0$, and
any positive number of initial inventions.  Hence no phase transition
occurs.  For small $p$, the population exhibits a long
bottleneck, where it grows slowly with time.  But after this quiescent
period, surviving populations explode faster than exponentially.   
Indeed both human populations~\cite{population} and technological advances (such
as computing speed for fixed cost~\cite{increasing}) are known to 
exhibit a `law of accelerating return'~\cite{accelerating} --
where growth occurs faster than exponentially. We
associate the sudden proliferation of innovations following a landmark
invention with such a super-explosion, rather than with a phase
transition.  We derive analytic results, including exact scaling laws
in the limit $p \to 0$, for the bottleneck and
transition to super-explosive growth that are confirmed by numerical
simulations.

At its most basic level, innovation can be represented as a growing
phylogenetic network. Phylogenetic networks~\cite{Moret04} are
generalizations of trees to the case of more than one ancestor. Such
trees find wide uses to depict evolutionary relationships between
e.g. genes or species, cultures~\cite{Fujikawa1996}, or
languages~\cite{gray::2003}. Each node in the network represents an element in the
population of innovations. It has one or more parent nodes, and can
branch to produce daughter nodes (new innovations) at future
generations.

Independent branching of each node corresponds to the Galton-Watson
(GW) branching process.  Assume that starting with a single root node,
each node on the tree independently produces a number of daughter
nodes that is Poisson distributed with mean $\mu$.  The whole tree
goes extinct only if each of the subtrees that follow the root's
daughters dies. Hence, the survival probability ${\cal Z}$ satisfies
$1-{\cal Z} = e^{-\mu {\cal Z}}$, with a non-zero solution only when
$\mu>1$. This phase transition that the GW process exhibits at $\mu=1$
is a general property of growth processes with independent branching.

In contrast, we recognize that innovation is a historically contingent
process driven by interactions.  Hence we put forward a more relevant
process -- the {\em interacting branching process} (IBP), as a
prototypical model.  The total population at generation $g$ is
$N_g$. It consists of survivors from previous generations and new
nodes created in generation $g$. At update $g+1$, if $S_{g+1}$
``young'' nodes are born, and $K_{g+1}$ ``old'' nodes are killed, 
\begin{equation}
 N_{g+1}=N_g+S_{g+1}-K_{g+1} \; .
\end{equation} 
Birth happens because each of the $S_g$ nodes in generation $g$ can
mate with each of the $N_g$ nodes currently alive (including young and
old ones, and even itself). Each mating pair produces a Poisson distributed number of
offspring with mean $p$. This makes the total number of offspring of a
node in a generation $g$ Poisson distributed with mean $U_g = p N_g$.
Hence the size of the next generation $S_{g+1}$ (given $S_g$ and $N_g$) is also Poisson
distributed, with mean ${\bar S}_{g+1} = p N_g S_g$.  Note that
$U_g$ in the IBP depends on $N_g$, unlike the GW process.

Killing nodes incorporates the possibility that inventions become
obsolete.  Killing also happens stochastically: After creating the new
$(g+1)^{th}$ generation, we kill and remove each of the $N_g$ old
individuals, with probability $p/\tau$. Hence the average $\bar
K_{g+1} = (p/\tau) N_g$.

In a mean field analysis, the stochastic variables $S_g$ and $K_g$ are
replaced by their average values ${\bar S}_g$ and ${\bar K}_g$. If
$\tau = p$ no nodes from older generations survive. Then $N_g=S_g$ for
all $g$ and ${\bar S}_{g+1} = p {\bar S}_g^2$. If the initial value
$S_0< 1/p$, the process always dies out, while it escapes extinction
if $S_0 > 1/p$ -- indicating a phase transition. However, due to
fluctuations, some populations can survive for any finite $p$, even
when $S_0=1$. The probability to survive goes to zero when $p\to 0$,
but it is non-zero for any $p > 0$.  In addition, there exists a critical
value of $\tau$ such that for $\tau> \tau_c$ the survival probability 
of the IBP has a precisely known essential singularity as $p \to 0$. We 
also find that fluctuations drive super-explosions for $\tau \leq \tau_c$; 
while for $\tau > \tau_c$ fluctuations are irrelevant. The precise 
value of $\tau_c$ depends on $S_0$ ($\tau_c \simeq 0.391$ for $S_0=1$), 
but it is finite as long as $S_0$ is finite. For brevity, we restrict 
$S_0=1$ in what follows.

For $\tau=\infty$ innovations never die, $K_g=0$ for all $g$, and
$N_g= \sum_{g'=0}^{g}S_{g'}$. Before the occurence of an empty
generation the branching ratio $U_g \geq p g$ is an increasing
function of $g$ with average slope $\geq p$. Hence, for some long
lived processes $U_g$ must exceed one at finite $g$. Comparison with
the GW process shows that after $U_g$ has irreversibly exceeded unity,
the population has a non-zero probability to escape extinction.
Notice that this is a rigorous result and contradicts the conclusions
of~\cite{Hanel05,Hanel07}.  Unlike the GW process, the average IBP
population size does not explode exponentially with time. It
super-explodes: Soon after $U_g$ exceeds one, the population size
grows faster than exponentially. In fact this is the almost certain
fate of an IBP that starts with $S_0 = N_0 \gg1/p$. Then we can ignore
fluctuations and set $S_{g+1}=p N_g S_g$, or
\begin{eqnarray*}
p S_{g+1} &=&p( p N_g S_g) = \left(\sum_{g'=0}^{g} p S_{g'}\right) p S_g \geq (p S_g)^2,
\end{eqnarray*}
indicating that $p S_g$ and hence $S_g$ grows faster than
$(pS_0)^{2^g}$ -- or faster than exponentially. We say that the
IBP reaches the super-explosive phase when $S_g$ increases
faster than exponentially with $g$, and now study how the IBP approaches
this regime.

For any $\tau$, if a population has survived to  generation $g$, it dies at the
$(g+1)^{th}$ generation if none of the $S_g$ nodes have any offspring,
an event that happens with probability $e^{-p N_g S_g}$.  Thus if
$Z(g)$ is the probability to survive to the $g^{th}$
generation, $Z(g+1)=\left( 1 -\langle e^{-p N_g S_g}
  \rangle'\right) Z(g)$.  The prime indicates that the average
is restricted to populations that survive to $g$. 
These are ``conditioned-to-grow'' populations, which do not include empty generations. Iterating this expression gives
\begin{eqnarray}
  \label{eq:VSsurvival}
   Z(g) &=& \prod_{g'=0}^{g-1}( 1 - \left \langle e^{-p N_{g'} S_{g'}}\right \rangle')\quad .
\end{eqnarray}
Setting $g = \infty$, we get the probability that the process escapes
extinction, ${\cal Z} \equiv Z(\infty)$.

For $p \ll 1$, almost all populations die out after a few generations,
frustrating a direct numerical approach to obtain $Z$. However,
Eq.~(\ref{eq:VSsurvival}) motivates a computationally efficient method
that {\it conditions} on surviving populations and ignores those that
die out. The conditioned-to-grow populations entering the expectation
value in Eq.~(\ref{eq:VSsurvival}) have a Poisson offspring distribution
\begin{eqnarray}
  \label{eq:truncatedPoisson}
  {\rm Prob}'[S_{g+1} = m|N_gS_g] &=&\frac{e^{-p N_g S_g }}{1-e^{-p  N_g S_g }} \frac{(p N_g S_g )^m}{m!} ,
\end{eqnarray}
for $m \geq 1$.  The trunctation at $m=1$ corresponds to conditioning
on survival. We record $N_g, S_g$ for each generation to obtain the death
probability $\langle e^{-p N_g S_g}\rangle'$.  We now define a
rescaled time $t\equiv pg$, and write the death
probability as $\langle e^{-p S_g N_g} \rangle' \equiv e^{-t}{\cal
  D}(t)$, in which case the logarithm of the survival probability $Z(t/p)$ obeys
\begin{eqnarray}
\label{eq:SPscaling}
\lim_{p \rightarrow 0} p ~{\rm log}~ Z(t/p) &=&\lim_{p \rightarrow 0} p \sum_{g'=0}^{t/p}{\rm ln}\langle 1 - e^{- p S_{g'} N_{g'}}\rangle'\nonumber\\
&=& \int_0^{t} d t' {\rm ln}\left( 1- e^{-t'}{\cal D}(t')\right) ,
\end{eqnarray} 
where the last expression is manifestly independent of $p$.

\begin{figure}
  \centering
  \includegraphics[width=8cm]{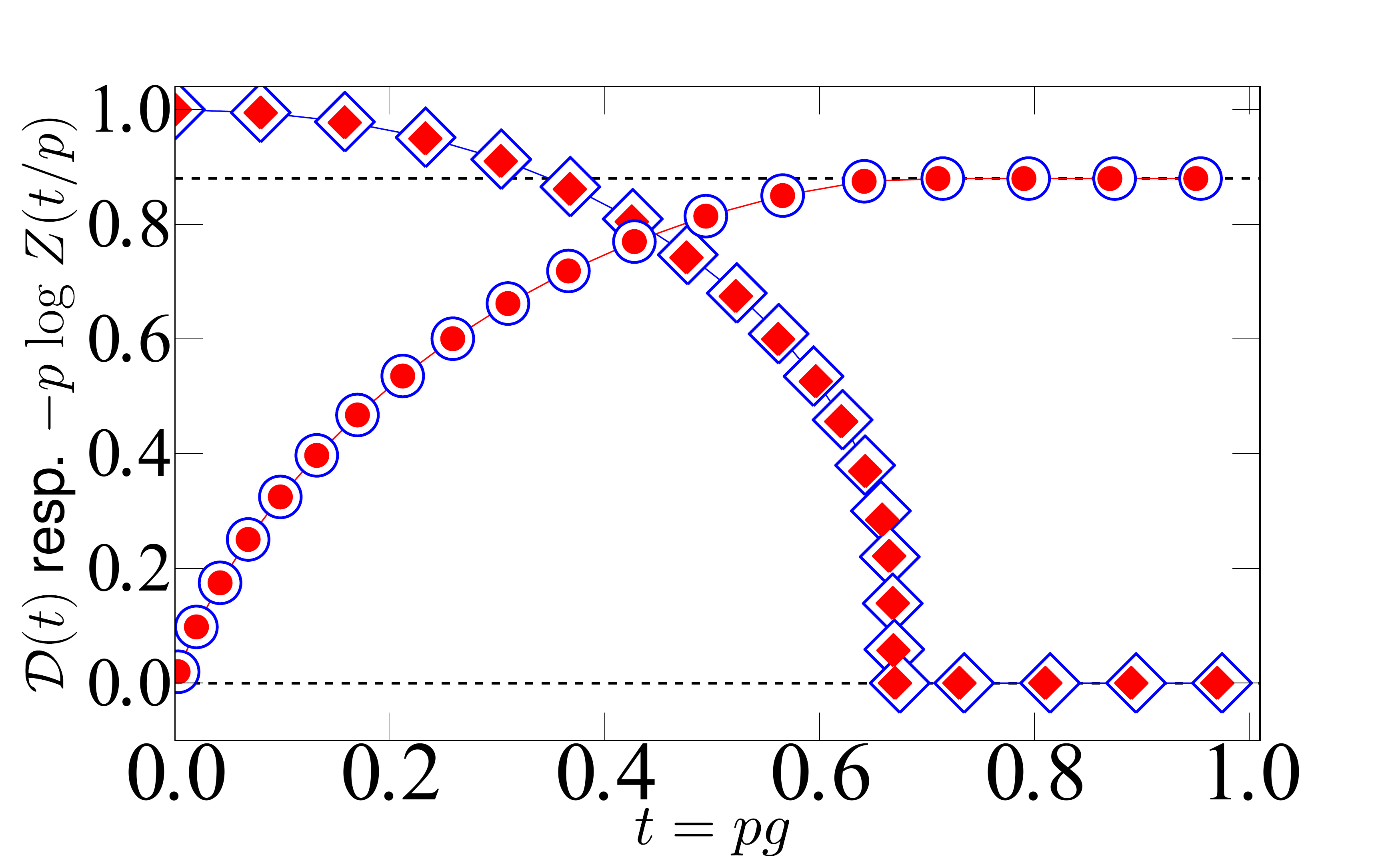}
  \caption{\label{fig:DPSPTlinfty}{ \bf Death and survival
    probabilities.} (color online) Diamonds: rescaled rate $ {\cal D}(t)$
    at which surviving populations die {\it vs.} scaled time $t = p g$,
    for two different values of $p\ll 1$, $ 10^{-6} (blue),$ and $10^{-7}
    (red)$. As $p \to 0$, the death probability converges to a limiting
    function that vanishes at $t = t_e \approx 0.67$. Circles: $-p\;
    \log Z(t/p)$ {\it vs.} $t$, for the same two values of $p$.}
\end{figure}

Fig.~\ref{fig:DPSPTlinfty} shows the scaled logarithm of the survival
probability as a function of rescaled time $t$, for $\tau=\infty$. It exhibits near-perfect 
collapse for different $p\ll 1$, in agreement with Eq.~(\ref{eq:SPscaling}). 
Fig.~\ref{fig:DPSPTlinfty} also indicates that the scaled death probability 
goes to zero at  $t =t_e \approx 0.67$ for $p\rightarrow 0$, 
and concomitantly the survival probability becomes constant for all subsequent 
times. Thus ${\cal Z}\sim e^{-c/p}$ for $p\to 0$, with $c\approx 0.89$. These 
numerical results support our claim that for $\tau=\infty$ the IBP has a 
finite probability to escape extinction for any $p\neq 0$.

We now show how the branching ratio $U_g=pN_g$ approaches unity from below, before
the IBP super-explodes.  Multiplying 
Eq.~(\ref{eq:truncatedPoisson}) by ${\rm Prob}'[S_g|N_g]$ and summing
over $S_g$ gives ${\rm Prob}'[S_{g+1}=m|N_g]$ conditioned on growth  for $m\geq 1$ and
fixed $N_g$. Using continuous time $t$  and changing notation, 
Eq.~(\ref{eq:truncatedPoisson})  leads to
\begin{equation}
\label{eq:genSizeDistFP}
P_{U(t)}(S, t+dt) =\sum_{S'} P_{U(t)}(S',t) \frac{e^{- S' U(t)}}{1-e^{-S' U(t)}} \frac{ (S' U(t))^{S}}{S!}.
\nonumber
\end{equation}
Setting the conditioning variable $U(t)=U$, numerical iterations of this equation
quickly converge to a stationary distribution $P_U(S)$ for any $U<1$.
Since $U(t)$ involves an integral of $S(t)$, it has both lower
fluctuations and slower variations compared to $S$, as long as
$U(t)<1$.  Hence, by the law of large numbers, $U(t)$ can be replaced by its mean ${\bar U}'(t)$ over
different realizations surviving to time $t$.  Then the
distribution of $S$ for a given $u\equiv {\bar U}'$ is obtained from the stationary
solution of
\begin{eqnarray}
\label{eq:genSizeStationary}
P_u(S) &=& \sum_{S'} P_u(S') \frac{e^{- S' u}}{1-e^{-S' u}} \frac{ (S' u)^{S}}{S!}.
\end{eqnarray}
Noting that as neither $p$, time, nor the age of the nodes enter
Eq.~(\ref{eq:genSizeDistFP}), we expect it to be valid for all $p\neq
0$, and also for all $\tau$ -- as long as $u<1$. 

\begin{figure}
  \centering
  \includegraphics[width=8cm]{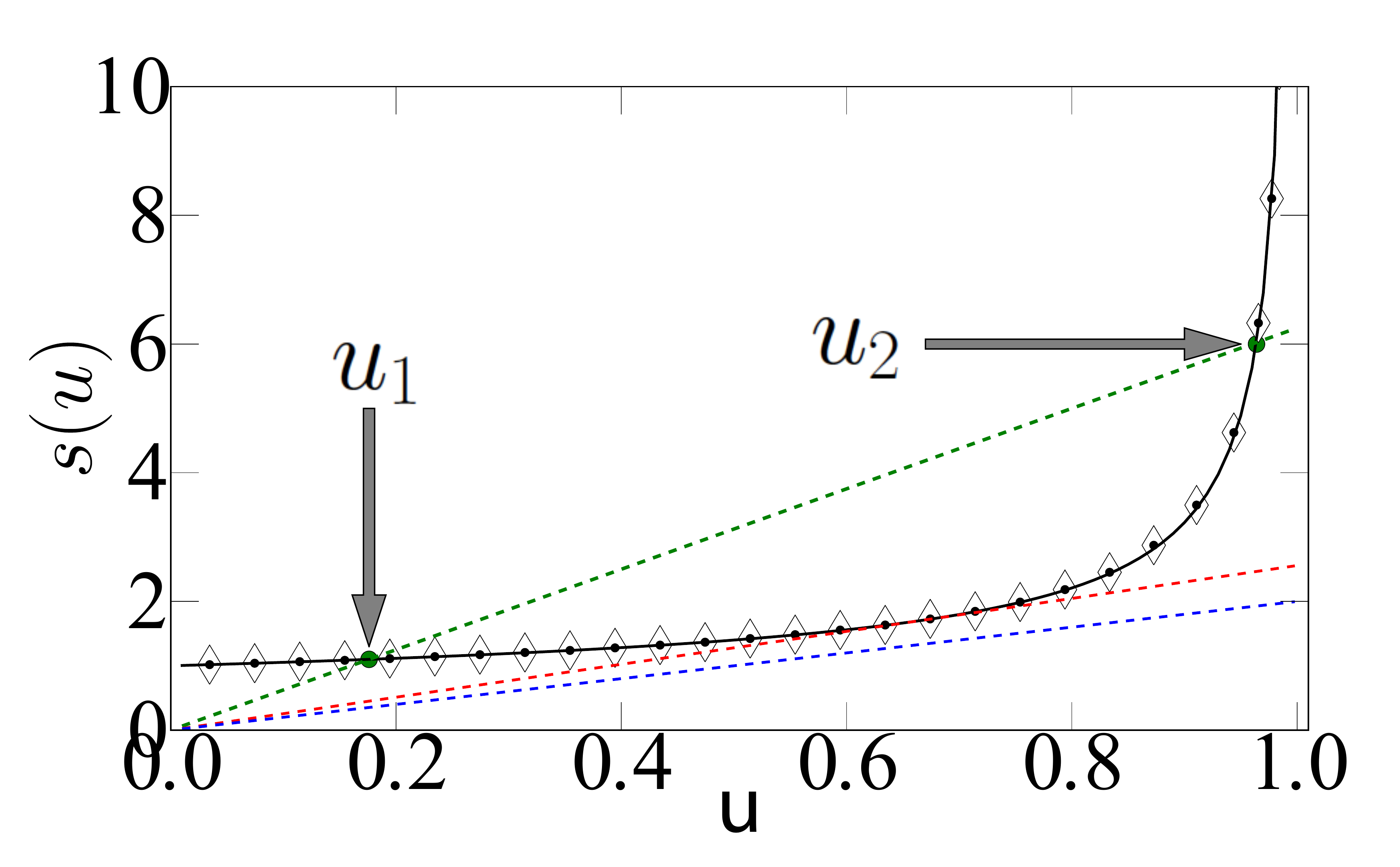}
  \caption{\label{fig:stationaryS}{\bf Mean generation size as a
function of branching ratio.} (color online) The solid black curve is
the numerical prediction for the mean generation size $s$ {\it vs.}
the mean branching ratio $u$ (both conditioned on growth) as obtained
from Eq.~(\ref{eq:genSizeStationary}). The diamonds are results of
numerical simulations of the IBP.  We used $p = 10^{-6}$ and
$\tau=\infty$, and averaged over $10^5$ conditioned-to-grow
populations. Agreement is excellent. Also plotted are three lines
showing $u/\tau$ on the y-axis, for three values of $\tau$ from top to
bottom: (1) 0.16, sub-critical $\tau<\tau_c$, (2) critical $\tau =
\tau_c =0.391$, and (3) 0.5, super-critical $\tau>\tau_c$,
respectively.}
\end{figure}


We have tested the distribution given by 
Eq.~(\ref{eq:genSizeStationary}) against averages over $10^5$
realizations of conditioned-to-grow IBP populations for small values of
$p$.  Fig. ~\ref{fig:stationaryS} shows our result
for the mean $s \equiv {\bar S}' = \sum_{S'} S' P_u(S')$ {\it vs.} $u$. 
The agreement is excellent not only
for the example $\tau = \infty$ shown, but also for all other
$\tau>\tau_c\simeq 0.391$.  The latter condition is explained
next.

For finite $\tau$, the branching ratio evolves as $\dot{U} =
S(t)-K(t)$, where $K(t)$ is the number of nodes killed at time $t$.
Neglecting fluctuations gives $\dot u= v(u)$ where $v(u)=s(u)
-u/\tau$. Fig.~\ref{fig:stationaryS} shows that $v(u)>0$ for all $u$,
if $\tau > \tau_c \approx 0.391$. As a result, $u(t)$ increases and
eventually super-explodes, so populations have a finite probability to
escape extinction. 

For $\tau < \tau_c$, $u/\tau$ intersects $s(u)$ at two values of $u$:
$u_1$ and $u_2$.  The smaller of these, $u_1$, gives the maximal value
of $u(t)$ reached before the process goes extinct. Including
fluctuations in a Langevin approach,  $\dot u= v(u)
+p^{\frac{1}{2}}\omega(u)\hat{\zeta}$, with $\hat{\zeta}$ a standard
Gaussian noise~\cite{noise}. Now the state with $u=u_1$ is
metastable: For any finite $p$, a finite fluctuation can kick the
system over the potential barrier to the unstable value $u_2$. Beyond
$u_2$, $s$ is larger than $u/\tau$ for all $u$, and surviving processes again
super-explode.
\begin{figure}
  \centering
  \includegraphics[width=8cm]{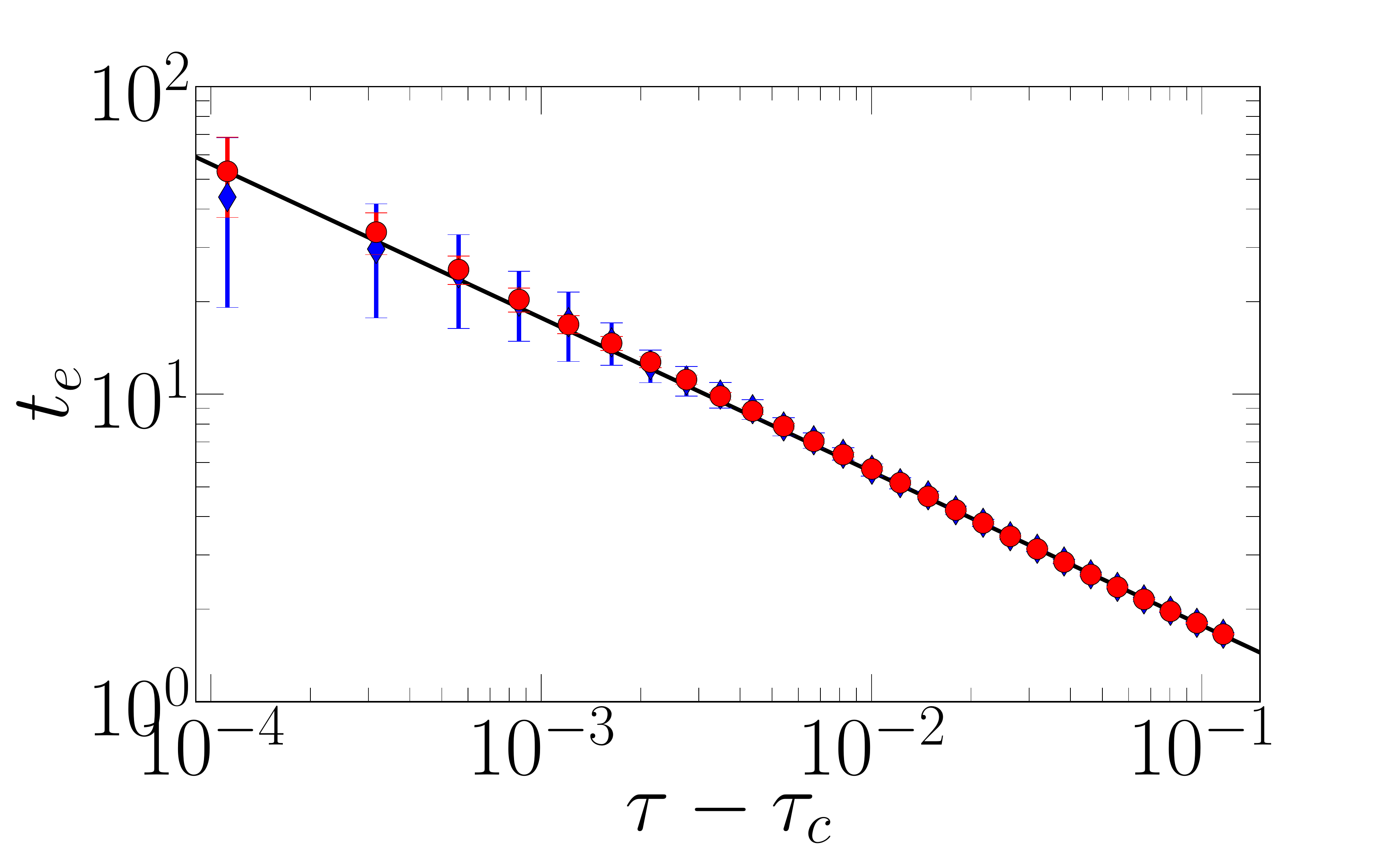}
  \caption{\label{fig:timetoenter}{\bf Mean time $\boldsymbol{t_e}$ to
enter the super-explosive phase}. (color online) The red circles (blue
diamonds ) are the results of
numerical simulations at $p=10^{-6}$ $(10^{-5})$ for the mean time to enter the
super-explosive phase, the black curve going through them a fit
$\propto(\tau -\tau_c)^{-1/2}$. }
\end{figure}

This scenario is supported by results of numerical simulations, which we
present in Fig.~\ref{fig:timetoenter}. 
The time, $t_e$, to super-explode
diverges as $(\tau- \tau_c)^{-1/2}$. It is independent of $p$, for $\tau>\tau_c$
provided $(\tau-\tau_c)$ is sufficiently large.  This
divergence is obtained analytically by expanding
$u$ about $u_{min}$ -- the value where $v(u)$ reaches its minimum
$v_{min}$. Writing $u=u_{min} + \epsilon$, one gets
$d{\tilde\epsilon}/d{\tilde t}= 1 + {\tilde \epsilon}^2/2 + o({\tilde
  \epsilon}^2)$. Here, ${\tilde \epsilon}=\epsilon\sqrt{s_{min}''/v_{min}}$,
${\tilde t}= t\sqrt{s_{min}''v_{min}}$, and $s_{min}''\equiv
d^2s/du^2|_{u=u_{min}}$. Surviving populations spend most of the time
around $u_{min}$ before reaching $u=1$ and super-exploding.  Hence, the
time to explode is $t_e \sim (s_{min}''v_{min})^{-1/2}$.  At $\tau=\tau_c$,
$v_{min}=0$.  A Taylor expansion near $\tau_c$ gives $v_{min}\sim \tau
-\tau_c$, and hence $t_e \sim (\tau -\tau_c)^{-1/2}$, in agreement
with Fig.~\ref{fig:timetoenter}. For finite $p$, the scaling breaks
down when the drift term $v_{min}$ becomes comparable to the noise, or
when $v_{min}\sim(\tau -\tau_c)\sim p^{1/2}$.

For $\tau < \tau_c$, the mean time to enter the explosive
phase, $t_e$, is no longer independent of $p$ in the limit
$p\rightarrow 0$.  For {\it e.g.} $\tau=p$, one has $U_g=pS_g$, and $U>1$
requires $S_g>1/p$.  The next paragraph shows that the average time to
first reach $S_g\geq 1/p$ is $t_e\sim \Gamma(1/p)p^{2 -(1/p)}$.

\begin{figure}
  \centering
  \includegraphics[width=8cm]{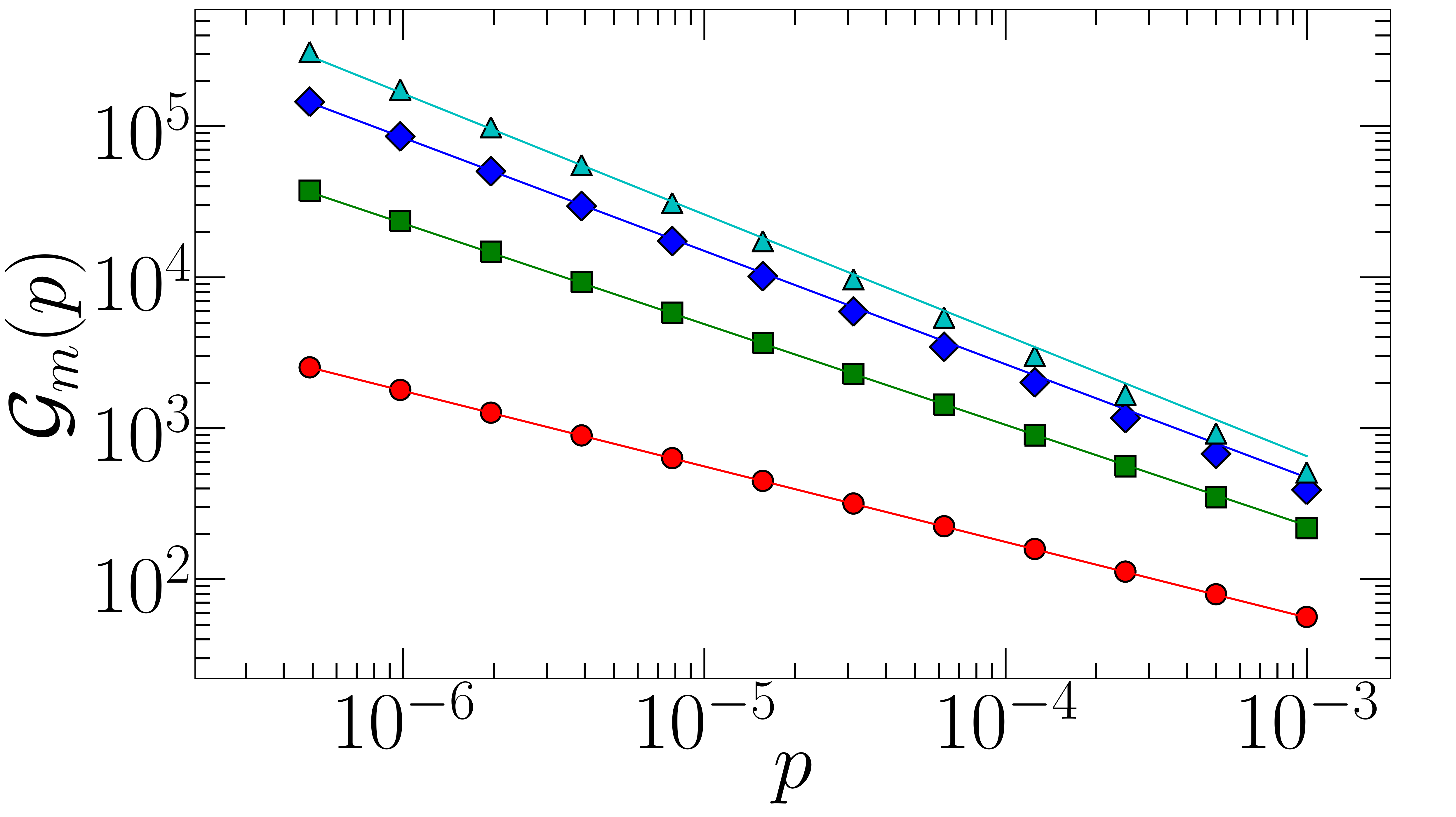}
  \caption{\label{fig:mfpt}{\bf Mean first passage (generation) times
      (MFPT) to reach a generation of size ${\boldsymbol m}$, for
      $\boldsymbol{\tau=\infty}$.} MFPTs for $m=2,3,4,5$ are shown from bottom to
    top respectively, along with the predicted values from
    Eq.~(\ref{mfpt}). For the range of $p$ studied, numerical
    results obtained by averaging over $10^6$ surviving processes agree with 
    predictions for $m=2,3$. For $m=4,5$ they converge to the predicted
    values as $p\to 0$.}
\end{figure}

Finally we estimate the mean first passage (rescaled) time,  ${\cal  T}_m(p) \equiv p {\cal G}_m(p)$, to a 
generation of size $m$ or larger in populations conditioned to reach such a generation size.
We derive  upper bounds by considering only the
single most probable path of evolution, which become exact in
the limit $p\to 0$. For any $\tau$ and small $p$, the
most likely surviving process before super-exploding is a simple {\it
chain} where $S_g=1$ for all $g$. For populations conditioned to reach $S_g\geq m$ for some $g$, the most likely
shape is a chain up to $g-1$, and a fan-out from $S_{g-1}=1$ to
$S_g=m$ during the last generation.  All other shapes are reduced by factors of $p$. Hence for $\tau = p$ (so $U_{g'} = p$
for all $g'<g$ in the chain) this gives the same relative chance
$p^{m-1}/m!$ to reach $S_g \geq m$ ( compared to $S_g=1$ ) at any
$g$. The probability $\xi_g^m$ that $S_g \geq m$ has not been reached evolves as $\xi_{g+1}^m = \left(1-p^{m-1}/m!\right)\xi_g^m$, and thus
\be {\cal G}_m(p) = \frac{{\cal T}_m(p)}{p} \sim m!\; p^{1-m} \quad ({\rm for}\;\; \tau =
p)\;.  \ee 
The time $t_e$ can be obtained by setting $m=1/p$.

For $\tau = \infty$, $U_{g} \approx pg$ is no longer independent of
$g$. The relative chance to reach $S_g \geq m$ ( compared to $S_g=1$ )
is $(pg)^{m-1}/m!$, and $\xi_{g+1}^m \approx (1-(pg)^{m-1}/m!)\xi_g^m$. 
Integrating over $g$ reveals the $g$-dependence of $\xi_g^m$, and that
\be {\cal G}_m(p) \propto
\left(\frac{1}{p}\right)^{1-1/m} \quad({\rm for}\;\;
\tau = \infty)\;.
                     \label{mfpt}
\ee
This is compared in Fig.~\ref{fig:mfpt} to direct simulations of the
IBP. It describes
the numerical results perfectly for $p\to 0$, and gives upper bounds for 
finite $p$, as it should. Eq.~(\ref{mfpt}) clarifies how surviving 
populations evolve in the limit $p\to 0$. First,
only one individual is born in each  generation. Generations
of size two  start appearing after $p^{-1/2}$ generations,
followed by the first appearance of a generation of size three after
$p^{-2/3}$ generations and so on. Finally after $1/p$ generations, 
generations of size $s\to\infty$ appear, indicating the onset of
super-explosive growth.

We have described autocatalytic networks of innovation in terms of an
interacting branching process (IBP). In contrast to standard branching
processes, two parents are needed to generate offspring. The IBP is
both analytically and numerically tractable. In mean field theory, it
shows a phase transition, which disappears due to fluctuations in a
rigorous microscopic treatment. When the probability $p$ to make new
innovations from any two existing ones is vanishingly small, we find
universal behavior that is independent of $p$. A super-explosive
phase, where the rate of new inventions grows faster than
exponentially follows a long quiescent bottleneck for $p\ll
1$.  This dynamics resembles the Dark Ages preceding
the Renaissance or the quiescent times between bursts of speciation in
punctuated biological evolution.  We speculate that our models unfolds large scale properties
of any co-evolutionary dynamics.  Indeed its super-explosive behavior is consistent with the law of accelerating returns
found in technological progress~\cite{increasing, accelerating}.


\begin{thebibliography}{99}
\bibitem{noise} The noise intensity $\omega(u) = \sqrt{s(u)(s(u)-1)+u/\tau}$ is independent of $p$.
\end{thebibliography}
\end{document}